\documentclass[jou,apacite]{apa6}

\usepackage[english]{babel}
\usepackage[utf8x]{inputenc}

\usepackage{times}
\usepackage{latexsym}
\usepackage{amsmath}
\usepackage{tikz}
\usepackage{url}
\usetikzlibrary{shapes}
\usetikzlibrary{fit,positioning,calc,backgrounds}
\usepackage{array,multirow,graphicx}

\usepackage[]{algorithm2e}
\usepackage{booktabs,caption}
\usepackage{bm}
\usepackage{amsfonts}
\usepackage{amssymb}
\usepackage{rotating}
\usepackage{array}
\newcolumntype{L}[1]{>{\raggedright\let\newline\\\arraybackslash\hspace{0pt}}m{#1}}
\newcolumntype{C}[1]{>{\centering\let\newline\\\arraybackslash\hspace{0pt}}m{#1}}
\newcolumntype{R}[1]{>{\raggedleft\let\newline\\\arraybackslash\hspace{0pt}}m{#1}}

\begin{document}


%


%

\title{Multitask Learning for Fine-Grained Twitter Sentiment Analysis }

\threeauthors{Georgios Balikas}{Simon Moura}{Massih-Reza Amini}
\threeaffiliations{Univ. Grenoble Alps, CNRS, Grenoble INP - LIG/Coffreo \\ geompalik@hotmail.com }{Univ. Grenoble Alps, CNRS, Grenoble INP - LIG \\ simon.moura@univ-grenoble-alpes.fr}{Univ. Grenoble Alps, CNRS, Grenoble INP - LJK \\ massih-reza.amini@univ-grenoble-alpes.fr}

%
%
%

%

\abstract{


Traditional sentiment analysis approaches tackle problems like ternary (3-category) and fine-grained (5-category) classification by learning the tasks separately. 
We argue that such classification tasks are correlated and we propose a multitask approach based on a recurrent neural network that benefits by jointly learning them. 
Our study demonstrates the potential of multitask models on this type of problems and improves the state-of-the-art results in the fine-grained sentiment classification problem.
 }
\maketitle

\section{Introduction}

Automatic classification of sentiment has mainly focused on categorizing tweets in either two (binary sentiment analysis) or three (ternary sentiment analysis) categories \cite{GiachanouC16}. 
In this work we study the problem of fine-grained sentiment classification where tweets are classified according to a five-point scale ranging from \textit{VeryNegative} to \textit{VeryPositive}. To illustrate this, Table \ref{tbl:example} presents examples of tweets associated with each of these categories. Five-point scales are widely adopted in review sites like Amazon and TripAdvisor,  where a user's sentiment is ordered with respect to its intensity. From a sentiment analysis perspective, this defines a classification problem with five categories. In particular,  Sebastiani et al. \cite{MartinoG016} defined such classification problems whose categories are explicitly ordered to be ordinal classification problems. To account for the ordering of the categories, learners are penalized according to how far from the true class their predictions are.

Although considering different scales, the various settings of sentiment classification are related. First, one may use the same feature extraction and engineering approaches to represent the text spans such as word membership in lexicons, morpho-syntactic statistics like punctuation or elongated word counts  \cite{balikasA16,kiritchenko2014sentiment}.
Second, one would expect that knowledge from one task can be transfered to the others and  this would benefit the performance.
Knowing that a tweet is ``Positive'' in the ternary setting narrows the classification decision between the \textit{VeryPositive} and \textit{Positive} categories in the fine-grained setting. From a research perspective this raises the question of whether and how one may benefit when tackling such related tasks and how one can transfer knowledge from one task to another during the training phase.

Our focus in this work is to exploit the relation between the sentiment classification settings and demonstrate the benefits stemming from combining them. To this end, we propose to formulate the different classification problems as a multitask learning problem and jointly learn them. Multitask learning \cite{Caruana97} has shown great potential in various domains and its benefits have been empirically validated \cite{CollobertW08,Plank16,LiuQH16,LiuQH16EMNLP} using different types of data and learning approaches.
An important benefit of multitask learning is that it provides an elegant way to access resources developed for similar tasks. By jointly learning correlated tasks, the amount of usable data increases. For instance, while for ternary classification one can label data using distant supervision with emoticons \cite{go2009twitter}, there is no straightforward way to do so for the fine-grained problem. However, the latter can benefit indirectly, if the ternary and fine-grained tasks are learned jointly.

The research question that the paper attempts to answer is the following: Can twitter sentiment classification problems, and fine-grained sentiment classification in particular, benefit from multitask learning? To answer the question, the paper brings the following two main contributions: (i) we show how jointly learning the ternary and fine-grained sentiment classification problems in a multitask setting improves the state-of-the-art performance,\footnote{An open implementation of the system for research purposes is available at \url{https://github.com/balikasg/sigir2017}.} and (ii) we demonstrate that recurrent neural networks outperform models previously proposed without access to huge corpora while being flexible to incorporate different sources of data.

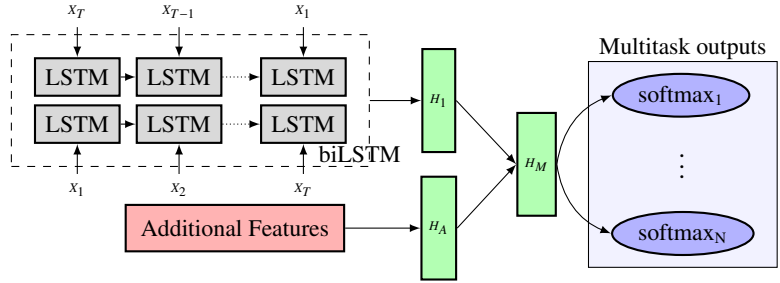
\begin{figure*}\centering
	\begin{tikzpicture}
		\tikzstyle{connect}=[draw, -latex]
		\node[rectangle, fill=gray!30, draw, thick, inner sep=4pt] (h1) {\small LSTM};
		\node[rectangle, fill=gray!30, draw, thick, inner sep=4pt, right = 0.2cm of h1] (h2) {\small LSTM};
		\node[rectangle, fill=gray!30, draw, thick, inner sep=4pt, right = 0.5cm of h2] (h3) {\small LSTM};
		
		\node[rectangle, fill=gray!30, draw, thick, inner sep=4pt, below=0.1cm of h1 ] (h1bis) {\small LSTM};
		\node[rectangle, fill=gray!30, draw, thick, inner sep=4pt, below=0.1cm of h2] (h2bis) {\small LSTM};
		\node[rectangle, fill=gray!30, draw, thick, inner sep=4pt, below=0.1cm of h3] (h3bis) {\small LSTM};
		
		\node[ above=0.4 of h1] (x1) {\tiny$X_{T}$};
		\node[ below=0.4 of h1bis] (x1bis) {\tiny$X_{1}$};

		\node[ above=0.4 of h2] (x2) {\tiny$X_{T-1}$};
		\node[ below=0.4 of h2bis] (x2bis) {\tiny$X_{2}$};
		
		\node[ above=0.4 of h3] (x3) {\tiny$X_{1}$};
		\node[ below=0.4 of h3bis] (x3bis) {\tiny$X_{T}$};

		\node[rectangle, fill=green!30, draw, thick, inner xsep=0.1cm, inner ysep=0.60cm, ] (hidden) at ($(h3)!0.5!(h3bis)+(1.8,0)$) {\tiny$H_1$};
		\node[rectangle, fill=green!30, draw, thick, inner xsep=0.1cm, inner ysep=0.60cm,  below  = 0.3cm  of hidden ] (hidden3)  {\tiny$H_A$};

		\node[rectangle, fill=green!30, draw, thick, inner xsep=0.1cm, inner ysep=0.60cm, ] (hidden2) at  ($(hidden3)!0.5!(hidden)+(1.3,0)$) {\tiny$H_M$};

		\node[rectangle, fill=red!30, draw, thick, inner xsep=0.20cm, inner ysep=0.20cm, left=1cm of hidden3] (additional)  {\small Additional Features};
		\node[ellipse, fill=blue!30, draw, thick, inner sep=2pt, above right = 0.01cm and 1.cm of hidden2] (o1) {\small softmax$_1$};
  		\node[ellipse, fill=blue!30, draw, thick, inner sep=2pt, below right = 0.01cm and 1.cm of hidden2] (o2) {\small softmax$_\text{N}$};
		\begin{scope}[on background layer]
			\node [draw,  fill=blue!5, fit= (o1) (o2), inner xsep=0.3cm, inner ysep=0.15cm, label={[xshift=0mm, yshift=-1mm]above : \small Multitask outputs}] {};
			\node [draw, dashed, fit= (h1) (h1bis) (h2) (h2bis) (h3) (h3bis), inner xsep=0.30cm, inner ysep=0.30cm, label={[xshift=7mm, yshift=4mm]below right: \small biLSTM}] (biLSTM) {};
		\end{scope}

		\path (h1.east) edge [connect]  (h2)
		(h1bis.east) edge [connect]  (h2bis)
		(h2) edge [densely dotted,connect] (h3)
		(h2bis) edge [densely dotted,connect] (h3bis)
		(biLSTM.east) edge [connect]  (hidden.west)
		(additional) edge [connect]  (hidden3)
		(hidden3.east) edge [connect] (hidden2.west)
		(hidden.east) edge [connect] (hidden2.west)
		(hidden2.east) edge [connect, bend left]  (o1.west)
		(hidden2.east) edge [connect, bend right]  (o2.west)
		(x1) edge [connect] (h1)
		(x2) edge [connect] (h2)
		(x3) edge [connect] (h3)
		(x1bis) edge [connect] (h1bis)
		(x2bis) edge [connect] (h2bis)
		(x3bis) edge [connect] (h3bis)
		(o1) -- node[auto=false,rotate=90]{\ldots} (o2);

	\end{tikzpicture}
	\caption{The neural network architecture for multitask learning. The biLSTM output is transformed by the hidden layers $H_1$, $H_M$ and is led to $N$ output layers, one for each of the tasks.  The lower part of the network can be used to incorporate additional information.}\label{fig:nn_architecture}

\end{figure*}

\begin{table}
\scriptsize\centering \setlength{\tabcolsep}{3pt}
	\begin{tabular}{l L{5.cm}}
		\toprule
		\texttt{VeryNegative} &     Beyond frustrated with my \#Xbox360 right now, and that as of June, @Microsoft doesn't support it. Gotta find someone else to fix the drive. \\[2pt]
		\texttt{Negative} &     @Microsoft Heard you are a software company. Why then is most of your software so bad that it has to be replaced by 3rd party apps?\\[2pt]
		\texttt{Neutral}  &     @ProfessorF @gilwuvsyou @Microsoft @LivioDeLaCruz We already knew the media march in ideological lockstep but it is nice of him to show it.\\[2pt]
		\texttt{Positive}  &     PAX Prime Thursday is overloaded for me with @Microsoft and Nintendo indie events going down. Also, cider!!! :p\\[2pt]
		\texttt{VeryPositive} &     I traveled to Redmond today. I'm visiting with @Microsoft @SQLServer engineers tomorrow - at their invitation. Feeling excited.\\[2pt]
		\bottomrule
	\end{tabular}
	\caption{The example demonstrates the different levels of sentiment a tweet may convey. Also, note the Twitter-specific use of language and symbols.}\label{tbl:example}
\end{table}

\section{Multitask Learning for Twitter Sentiment Classification}

In his work, Caruana \cite{Caruana97} proposed a multitask approach in which a learner takes advantage of the multiplicity of interdependent tasks while jointly learning them. 
The intuition is that if the tasks are correlated, the learner can learn a model jointly for them while taking into account the shared information which is expected to improve its generalization ability.
People express their opinions online on various subjects (events, products..), on several languages  and in several styles (tweets, paragraph-sized reviews..), and it is exactly this variety that motivates the multitask approaches. Specifically for Twitter for instance, the different settings of classification like binary, ternary and fine-grained are correlated since their difference lies in the sentiment granularity of the classes which increases while moving from binary to fine-grained problems.

There are two main decisions to be made in our approach: the learning algorithm, which learns a decision function, and the data representation.
With respect to the former, neural networks are particularly suitable as one can design architectures with different properties and arbitrary complexity. Also, as training neural network usually relies on back-propagation of errors, one can have shared parts of the network trained by estimating errors on the joint tasks and others specialized for particular tasks. Concerning the data representation, it strongly depends on the data type available. For the task of sentiment classification of tweets with neural networks, distributed embeddings of words have shown great potential. Embeddings are defined as low-dimensional, dense representations of words that can be obtained in an unsupervised fashion by training on large quantities of text \cite{PenningtonSM14}.


Concerning the neural network architecture, we focus on Recurrent Neural Networks (RNNs) that are capable of modeling short-range and long-range dependencies like those exhibited in sequence data of arbitrary length like text. While in the traditional information retrieval paradigm such dependencies are captured using $n$-grams and skip-grams, RNNs learn to capture them automatically \cite{DyerBLMS15}. To circumvent the problems with capturing long-range dependencies and preventing gradients from vanishing, the long short-term memory network (LSTM) was proposed \cite{hochreiter1997long}. In this work, we use an extended version of LSTM called bidirectional LSTM (biLSTM).  While standard LSTMs access information only from the past (previous words), biLSTMs capture both past and future information effectively \cite{HuangXY15,DyerBLMS15}. They consist of two LSTM networks, for propagating text forward and backwards with the goal being to capture the dependencies better. Indeed, previous work on multitask learning showed the effectiveness of biLSTMs in a variety of problems:  \cite{alonso2016multitask} tackled sequence prediction, while \cite{Plank16} and \cite{KiperwasserG16} used biLSTMs for Named Entity Recognition and dependency parsing respectively. 

Figure \ref{fig:nn_architecture} presents the architecture we use for multitask learning. In the top-left of the figure a biLSTM network (enclosed by the dashed line) is fed with embeddings $\{X_1, \ldots, X_T\}$ that correspond to the $T$ words of a tokenized tweet. Notice, as discussed above, the biLSTM consists of two LSTMs that are fed with the word sequence forward and backwards.  On top of the biLSTM network  one (or more) hidden layers $H_1$ transform its output. The output of $H_1$ is led to the softmax layers for the prediction step. There are $N$ softmax layers and each is used for one of the $N$  tasks of the multitask setting. In tasks such as sentiment classification, additional features like membership of words in sentiment lexicons or counts of elongated/capitalized words can be used to enrich the representation of tweets before the classification step \cite{kiritchenko2014sentiment}. The lower part of the network illustrates how such sources of information can be incorporated to the process. A vector ``Additional Features'' for each tweet is transformed from the hidden layer(s) $H_A$ and then is combined by concatenation with the transformed biLSTM output in the $H_M$ layer.

\section{Experimental setup}
Our goal is to demonstrate how multitask learning can be successfully applied on the task of sentiment classification of tweets. The particularities of tweets are to be short and informal text spans. The common use of abbreviations, creative language etc., makes the sentiment classification problem challenging. 
To validate our hypothesis, that learning the tasks jointly can benefit the performance, we propose an experimental setting where there are data from two different twitter sentiment classification problems: a fine-grained and a ternary. We consider the fine-grained task to be our primary task as it is more challenging and obtaining bigger datasets, \textit{e.g.} by distant supervision, is not straightforward and, hence we report the performance achieved for this task.

Ternary and fine-grained sentiment classification were part of the SemEval-2016 ``Sentiment Analysis in Twitter'' task \cite{NakovRRSS16}. We use the high-quality datasets the challenge organizers released.\footnote{The datasets are those of Subtasks A and C, available at \url{http://alt.qcri.org/semeval2016/task4/}.} The dataset for fine-grained classification is split in training, development, development\_test and test parts. In the rest, we refer to these splits as \texttt{train, development} and \texttt{test}, where \texttt{train} is composed by the training and the development instances. Table \ref{tbl:datasets} presents an overview of the data. As discussed in \cite{NakovRRSS16} and illustrated in the Table, the fine-grained dataset is highly unbalanced and skewed towards the positive sentiment: only $13.6\%$ of the training examples are labeled with one of the negative classes.

\noindent\textbf{Feature representation}
We report results using two different feature sets. The first one, dubbed \texttt{nbow}, is a  neural bag-of-words  that uses text embeddings to generate low-dimensional, dense representations of the tweets.  To construct the \texttt{nbow} representation, given the word embeddings dictionary where each word is associated with a vector, we apply the \texttt{average} compositional function that averages the embeddings of the words that compose a tweet. Simple compositional functions like  \texttt{average} were shown to be robust and efficient in previous work \cite{MitchellL10}. Instead of training embeddings from scratch, we use the pre-trained on tweets GloVe embeddings of \cite{PenningtonSM14}.\footnote{url{http://nlp.stanford.edu/data/glove.twitter.27B.zip}}
%
In terms of resources required, using only \texttt{nbow} is efficient as it does not require any domain knowledge. However, previous research on sentiment analysis showed that using extra resources, like sentiment lexicons, can benefit significantly the performance \cite{kiritchenko2014sentiment,balikasA16}.
To validate this and examine at which extent neural networks and multitask learning benefit from such features we evaluate the models using  an augmented version of \texttt{nbow}, dubbed \texttt{nbow+}. The feature space of the latter, is augmented using 1,368 extra features consisting mostly of counts of punctuation symbols ('!?\#@'), emoticons, elongated words and word membership features in several sentiment lexicons. Due to space limitations, for a complete presentation of these features, we refer the interested reader to \cite{balikasA16}, whose open implementation we used to extract them.\footnote{\url{https://github.com/balikasg/SemEval2016-Twitter_Sentiment_Evaluation}}

\begin{table}\scriptsize\centering
 \begin{tabular}{r c ccccc}
 \toprule
  & $|D|$ &  VeryNeg. & Neg. & Neutr. & Pos. & VeryPos. \\
  \midrule
 Train & 7,292  &111 & 884 & 2,019 & 3,726 & 432 \\
 Dev. & 1,778  & 29 & 204 &533 & 887 & 125 \\
 Test & 20,632 & 138 & 2,201 & 10,081 & 7,830 & 382 \\
 Ternary & 5,500 &- & 785 & 1,887 & 2,828 & - \\
 \bottomrule
 \end{tabular}
\caption{Cardinality and class distributions of the datasets. }\label{tbl:datasets}
\end{table}

\noindent\textbf{Evaluation measure}
To reproduce the setting of the SemEval challenges \cite{NakovRRSS16}, we optimize our systems using as primary measure the macro-averaged Mean Absolute Error ($MAE_M$) given by:
\begin{small}
\[
 MAE_M=\frac{1}{|C|}\sum\limits_{j=1}^{|C|} \frac{1}{|\text{Te}_j|}\sum\limits_{x_i\in\text{Te}_j} |h(x_i)-y_i|
\]
\end{small}
\noindent where $|C|$ is the number of categories, $\text{Te}_j$ is the set of instances whose true class is $c_j$, $y_i$ is the true label of the instance $x_i$ and $h(x_i)$ the predicted label. The measure penalizes decisions far from the true ones and is macro-averaged to account for the fact that the data are unbalanced. Complementary to $MAE_M$, we report the performance achieved on the micro-averaged $F_1$ measure, which is a commonly used measure for classification. 

\noindent\textbf{The models} To evaluate the multitask learning approach, we compared it with several other models. 
Support Vector Machines (SVMs) are maximum margin classification algorithms that have been shown to achieve competitive performance in several text classification problems \cite{NakovRRSS16}. \texttt{SVM$_{\text{ovr}}$} stands for an SVM with linear kernel and an one-vs-rest approach for the multi-class problem. Also, \texttt{SVM$_\text{cs}$} is an SVM with linear kernel that employs the crammer-singer strategy \cite{CrammerS01} for the multi-class problem. 
Logistic regression (LR) is another type of linear classification method, with probabilistic motivation.
Again, we use two types of Logistic Regression depending on the multi-class strategy: \texttt{LR$_\text{ovr}$} that uses an one-vs-rest approach and multinomial Logistic Regression  also known as the \texttt{MaxEnt} classifier that uses a multinomial criterion. 

Both SVMs and LRs as discussed above treat the problem as a multi-class one, without considering the ordering of the classes. 
For these four models, we tuned the hyper-parameter $C$ that controls the importance of the L$_2$ regularization part in the optimization problem  with grid-search over $\{10^{-4},\ldots, 10^4\}$  using 10-fold cross-validation in the union of the training and development data and then retrained the models with the selected values. 
Also, to account for the unbalanced classification problem we used class weights to penalize more the errors made on the rare classes. These weights were inversely proportional to the frequency of  each class.
For the four models we used the implementations of Scikit-learn \cite{scikit-learn}.

For multitask learning we use the architecture shown in Figure \ref{fig:nn_architecture}, which we implemented with Keras \cite{keras}. The embeddings are initialized with the 50-dimensional GloVe embeddings while the output of the biLSTM network is set to dimension 50. The activation function of the hidden layers is the hyperbolic tangent.
The weights of the layers were initialized from a uniform distribution, scaled as described in \cite{GlorotB10}.  We used the Root Mean Square Propagation optimization method.
We used dropout for regularizing the network. We trained the network using batches of 128 examples as follows: before selecting the batch, we perform a Bernoulli trial with probability $p_M$ to select the task to train for. With probability $p_M$ we pick a batch for the fine-grained sentiment classification problem, while with probability $1-p_M$ we pick a batch for the ternary problem. As shown in Figure \ref{fig:nn_architecture}, the error is backpropagated until the embeddings, that we fine-tune during the learning process. Notice also that the weights of the network until the layer $H_M$ are shared and therefore affected by both tasks.

To tune the neural network hyper-parameters we used 5-fold cross validation. We tuned the probability $p$ of dropout after the hidden layers $H_M, H_1, H_A$  and for the biLSTM for $p\in \{0.2, 0.3, 0.4, 0.5\}$, the size of the hidden layer $H_M \in \{20, 30, 40, 50\}$ and the probability $p_M$ of the Bernoulli trials from $\{0.5, 0.6, 0.7, 0.8\}$.\footnote{Overall, we cross-validated 512 combinations of parameters. The best parameters were: 0.2 for all dropout rates, 20 neurons for $H_M$ and $p_M=0.5$.} During training, we monitor the network's performance on the development set and apply early stopping if the performance on the 
validation set does not improve for 5 consecutive epochs.


\noindent\textbf{Experimental results} Table \ref{tbl:results} illustrates the performance of the models for the different data representations. The upper part of the Table summarizes the performance of the baselines. The entry ``Balikas et al.'' stands for the winning system of the 2016 edition of the challenge \cite{balikasA16}, which to the best of our knowledge holds the state-of-the-art. Due to the stochasticity of training the biLSTM models, we repeat the experiment 10 times and report the average and the standard deviation of the performance achieved. 

Several observations can  be made from the table. First notice that, overall, the  best performance is achieved by the neural network architecture that uses multitask learning. This entails that the system makes use of the available  resources efficiently and improves the state-of-the-art performance. In conjunction with the fact that we found the optimal probability $p_M=0.5$, this highlights the benefits of multitask learning over single task learning. Furthermore, as described above, the neural network-based models have only access to the training data as the development are hold for early stopping. On the other hand, the baseline systems were retrained on the union of the train and development sets. Hence, even with fewer resources available for training on the fine-grained problem, the neural networks outperform the baselines.
We also highlight the positive effect of the additional features that previous research proposed. Adding the features both in the baselines and in the biLSTM-based architectures improves the $MAE_M$ scores by several points. 



Lastly, we compare the performance of the baseline systems with the performance of the state-of-the-art system of \cite{balikasA16}. 
While  \cite{balikasA16} uses n-grams (and character-grams) with $n>1$, the baseline systems (SVMs, LRs) used in this work use the \texttt{nbow+} representation, that relies on unigrams. 
Although they perform on par, the competitive performance of \texttt{nbow} highlights the potential of distributed representations for short-text classification.
Further, incorporating structure and distributed  representations leads to the gains of the biLSTM network, in the multitask and single task setting.

\begin{figure}\centering
 \includegraphics[scale=0.55]{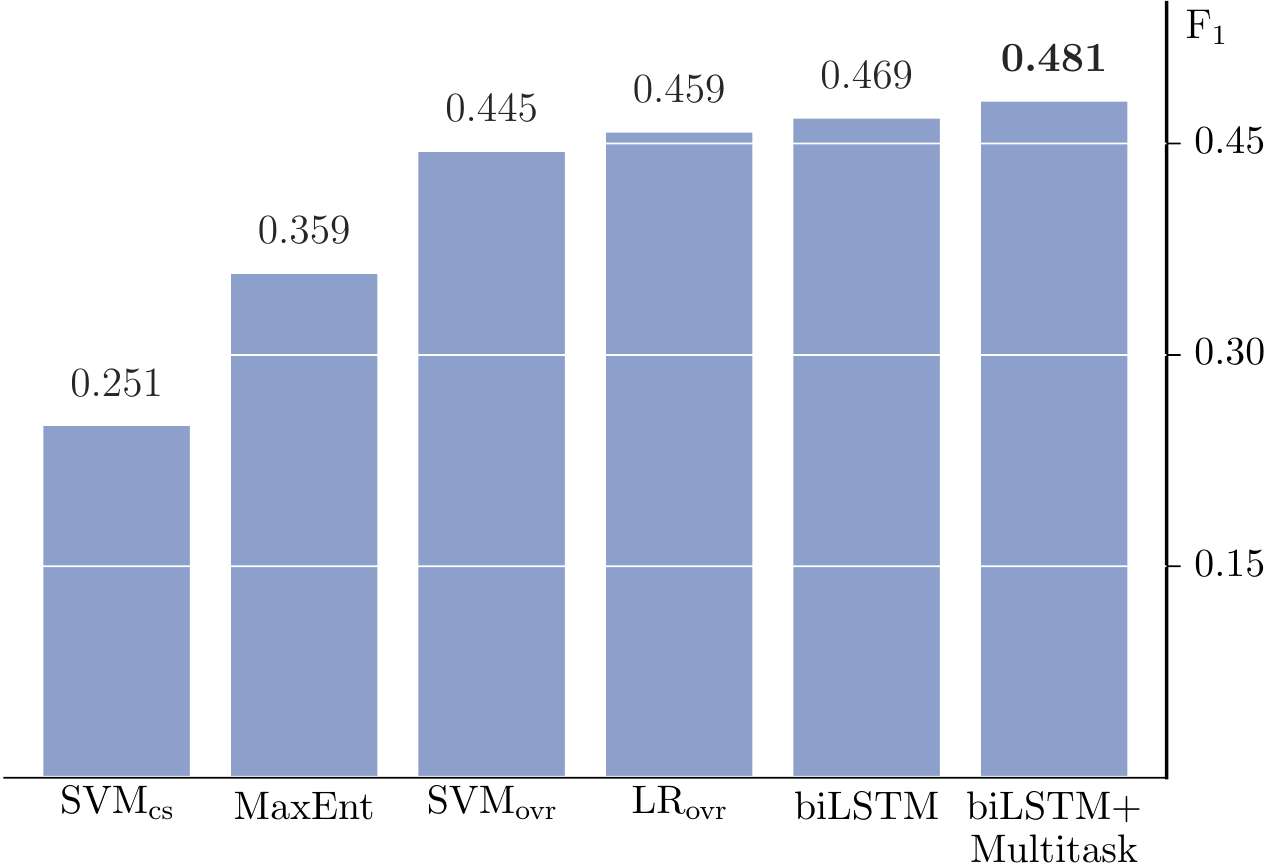}
 \caption{$F_1$ scores using the \texttt{nbow+} representations. The best performance is achieved with the multitask setting.}\label{fig:f1_measure}
\end{figure}

Similar observations can be drawn from Figure \ref{fig:f1_measure} that presents the $F_1$ scores. Again, the biLSTM network with multitask learning achieves the best performance. It is also to be noted that although the two evaluation measures are correlated in the sense that the ranking of the models is the same, small differences in the $MAE_M$ have large effect on the scores of the $F_1$ measure.

\begin{table}[t]
        \centering\small
        \begin{tabular}{l cc}
                \toprule
		    &  \texttt{nbow} &  \texttt{nbow+} \\\midrule
SVM$_{\text{ovr}}$  &  0.840 &  0.714 \\ 
SVM$_{\text{cs}}$   &  0.946 &  0.723 \\   
LR$_{\text{ovr}}$   &  0.836 &  0.712 \\
MaxEnt		    &  0.842 &  0.715 \\
\cite{balikasA16} & - & 0.719 \\
\midrule
biLSTM (single task)             &  0.827$\pm 0.017$  & 0.694$\pm 0.04$  \\
biLSTM+Multitask    & 0.786$\pm 0.025$ & \textbf{0.685}$\pm 0.024$  \\\bottomrule
        \end{tabular}
	\caption{The scores on $MAE_M$ for the systems. The best (lowest) score is shown in bold and is achieved in the multitask setting with the biLSTM architecture of Figure \ref{fig:nn_architecture}. }\label{tbl:results}
\end{table}

\section{Conclusion}
In this paper, we showed that by jointly learning the tasks of ternary and fine-grained classification with a multitask learning model, one can greatly improve the performance on the second. This opens several avenues for future research. Since sentiment is expressed in different textual types like tweets and paragraph-sized reviews, in different languages (English, German, ..) and in different granularity levels (binary, ternary,..) one can imagine multitask approaches that could benefit from combining such resources. Also, while we opted for biLSTM networks here, one could use convolutional neural networks or even try to combine different types of networks and tasks to investigate the performance effect of multitask learning.  Lastly, while our approach mainly relied on the foundations of \cite{Caruana97}, the internal mechanisms and the theoretical guarantees of multitask learning remain to be better understood.

\section{Acknowledgements} 
This work is partially supported by the CIFRE N 28/2015.

\bibliographystyle{ACM-Reference-Format}
\bibliography{references}  

\end{document}